\documentclass[conference]{IEEEtran} 
\ifCLASSOPTIONcompsoc
  \usepackage[nocompress]{cite}
\else
  \usepackage{cite}
\fi
\ifCLASSINFOpdf
\else
\fi
\usepackage[colorlinks=true,
            linkcolor=red,
            urlcolor=blue,
            citecolor=blue]{hyperref}
\usepackage{amsmath}
\usepackage{graphicx}
\usepackage{mathtools}
\usepackage{relsize}
\usepackage{graphicx}
\usepackage{algorithm}
\usepackage{filecontents,lipsum}
\usepackage{cite}
\usepackage{comment}
\usepackage{multirow}
\usepackage{amsfonts}
\usepackage[noend]{algpseudocode}
\usepackage[utf8]{inputenc}
\usepackage{graphicx}
\usepackage{caption}
\usepackage{latexsym}
\usepackage{etoolbox}
\usepackage{subcaption, float}
\makeatletter
\usepackage[inline]{enumitem}
\def\BState{\State\hskip-\ALG@thistlm}
\makeatother

\AfterEndEnvironment{table}{\vskip-1ex}
\begin{document}

%
\title{Transformative Influence of LLM and AI Tools in Student Social Media Engagement: Analyzing Personalization, Communication Efficiency, and Collaborative Learning}


\author{\IEEEauthorblockN{Masoud Bashiri\IEEEauthorrefmark{1} and Kamran Kowsari\IEEEauthorrefmark{1}}

\IEEEauthorblockA{\IEEEauthorrefmark{1} Department of System and Information Engineering,
University of Virginia,
Charlottesville, VA, USA}

\{ 
\href{mailto:mb4bw@virginia.edu}{mb4bw}, 
 \href{mailto:kk7nc@virginia.edu}{kk7nc}\}@virginia.edu}


%


\maketitle
\begin{abstract}
The advent of Large Language Models (LLMs) and Artificial Intelligence (AI) tools has revolutionized various facets of our lives, particularly in social media. These advancements have unlocked unprecedented opportunities for learning, collaboration, and personal growth for students. AI-driven applications transform how students interact with social media, offer personalized content and recommendations, and enable smarter, more efficient communication. Recent studies utilizing data from UniversityCube underscore the profound impact of AI tools on students' academic and social experiences. These studies reveal that students engaging with AI-enhanced social media platforms report higher academic performance, enhanced critical thinking skills, and increased engagement in collaborative projects.

Moreover, AI tools assist in filtering out distracting content, allowing students to concentrate more on educational materials and pertinent discussions. The integration of LLMs in social media has further facilitated improved peer-to-peer communication and mentorship opportunities. AI algorithms effectively match students based on shared academic interests and career goals, fostering a supportive and intellectually stimulating online community, thereby contributing to increased student satisfaction and retention rates.

In this article, we delve into the data provided by UniversityCube to explore how LLMs and AI tools are specifically transforming social media for students. Through case studies and statistical analyses, we offer a comprehensive understanding of the educational and social benefits these technologies offer. Our exploration highlights the potential of AI-driven tools to create a more enriched, efficient, and supportive educational environment for students in the digital age.\\
\end{abstract}

\begin{IEEEkeywords} 
LLM; Machine Learning; Social Network; UniversityCube
\end{IEEEkeywords}

\IEEEpeerreviewmaketitle

In today's digital age, the integration of technology in education has transformed the traditional classroom into a dynamic and interactive learning environment. Among the myriad technological innovations reshaping the educational landscape, educational social networks stand out as powerful platforms for facilitating collaboration, knowledge sharing, and personalized learning experiences among students. By harnessing the capabilities of Large Language Models (LLMs), AI-driven visualization tools, and other advanced technologies, these platforms offer unprecedented opportunities for enhancing student engagement, comprehension, and academic success.

Recent studies utilizing data from UniversityCube\footnote{\href{https://www.universitycube.net/}{https://www.universitycube.net/}}. have shown that AI tools significantly impact students' academic and social experiences. These studies reveal that students who engage with AI-enhanced social media platforms report higher levels of academic performance, improved critical thinking skills, and increased engagement in collaborative projects. Additionally, AI tools help in filtering out distracting content, allowing students to focus more on educational materials and relevant discussions.

Personalized chatbot-based teaching assistants can be crucial in addressing increasing classroom sizes, especially where direct teacher presence is limited. Large language models (LLMs) offer a promising avenue, with increasing research exploring their educational utility. However, the challenge lies not only in establishing the efficacy of LLMs but also in discerning the nuances of interaction between learners and these models, which impact learners' engagement and results. Harsh Kumar \textit {et. al.} \cite{kumar2023impact} conducted a formative study in an undergraduate computer science classroom and a controlled experiment on Prolific to explore the impact of four pedagogically informed guidance strategies on the learners' performance, confidence, and trust in LLMs. Direct LLM answers marginally improved performance, while refining student solutions fostered trust. Structured guidance reduced random queries as well as instances of students copy-pasting assignment questions to the LLM. Our work highlights the role that teachers can play in shaping LLM-supported learning environments.

Moreover, the integration of LLMs in social media has facilitated better peer-to-peer communication and mentorship opportunities. AI algorithms can match students with similar academic interests and career goals, fostering a more supportive and intellectually stimulating online community. This has led to a marked increase in student satisfaction and retention rates.

In this article, we will delve deeper into the data provided by UniversityCube to explore the specific ways in which LLM and AI tools are transforming social media for students. We will examine case studies and statistical analyses to provide a comprehensive understanding of the educational and social benefits these technologies offer. Through this exploration, we aim to highlight the potential of AI-driven tools to create a more enriched, efficient, and supportive educational environment for students in the digital age.\cite{kumar2023impact}. 
The convergence of educational social networks with LLMs, visualization tools, and AI-driven solutions represents a paradigm shift in how students access, interact with, and assimilate knowledge. From personalized content recommendations and real-time feedback on assignments to immersive visualizations of complex concepts and data, these platforms empower students to take charge of their learning journey and thrive in an increasingly interconnected and fast-paced world.

This paper explores the transformative role of educational social networks in leveraging LLMs, visualization tools, and AI-driven solutions to enhance learning outcomes and foster a culture of collaboration, innovation, and inclusivity in education. Through a comprehensive review of relevant literature, case studies, and empirical evidence, we examine the potential of these platforms to revolutionize the educational experience for students across diverse disciplines and learning environments.

Drawing insights from research findings, industry trends, and practical applications, we elucidate the key features, benefits, and challenges associated with the integration of LLMs, visualization tools, and AI-driven solutions in educational social networks. Furthermore, we explore the implications of these technologies for student engagement, academic performance, and well-being, highlighting best practices and emerging opportunities for future research and development.



%

\section{LLM in Education}
In today's rapidly evolving landscape of Artificial Intelligence, large language models (LLMs) have emerged as a vibrant research topic. LLMs find applications in various fields and contribute significantly. Despite their powerful language capabilities, similar to pre-trained language models (PLMs), LLMs still face challenges in remembering events, incorporating new information, and addressing domain-specific issues or hallucinations. To overcome these limitations, researchers have proposed Retrieval-Augmented Generation (RAG) techniques, and some have proposed the integration of LLMs with Knowledge Graphs (KGs) to provide factual context, thereby improving performance and delivering more accurate feedback to user queries\cite{latif2023knowledge,oduor2016kenyan,bui2024cross}.

Education plays a crucial role in human development and progress. With technological transformation, traditional education is being replaced by digital or blended education. Consequently, educational data in the digital environment is increasing day by day. Data in higher education institutions are diverse, comprising various sources such as unstructured/structured text, relational databases, web/app-based API access, etc. Constructing a Knowledge Graph from these cross-data sources is not a simple task. This article proposes a method for automatically constructing a Knowledge Graph from multiple data sources and discusses some initial applications (experimental trials) of KG in conjunction with LLMs for question-answering tasks\cite{latif2023knowledge,oduor2016kenyan,bui2024cross}.

Furthermore, this study proposes a method for knowledge distillation (KD) of fine-tuned Large Language Models (LLMs) into smaller, more efficient, and accurate neural networks. We specifically target the challenge of deploying these models on resource-constrained devices. Our methodology involves training the smaller student model (Neural Network) using the prediction probabilities (as soft labels) of the LLM, which serves as a teacher model. This is achieved through a specialized loss function tailored to learn from the LLM's output probabilities, ensuring that the student model closely mimics the teacher's performance\cite{latif2023knowledge,oduor2016kenyan,bui2024cross}.

\section{Use of LLM and AI Tools in Educational Social Networks}
The integration of Large Language Models (LLMs) and Artificial Intelligence (AI) tools in educational social networks represents a significant leap forward in how students engage with their academic environments. These advanced technologies offer numerous benefits that enhance learning experiences, foster collaboration, and support academic success\cite{meier2024llm}.

\subsection{Personalized Learning and Support}
One of the primary advantages of LLMs in educational social networks is the ability to provide personalized learning experiences based on the behavior\cite{kashikarladder}. AI algorithms can analyze a student's interaction patterns, performance data, and learning preferences to deliver customized content and recommendations\cite{laak2024ai}. This tailored approach ensures that students receive the specific resources and support they need to excel in their studies. For example, AI-driven platforms can suggest relevant articles, study materials, and practice exercises based on individual progress and areas that require improvement\cite{yang2024harnessing}.

UniversityCube data shows that students who receive personalized content recommendations from AI tools experience a significant improvement in their academic performance and satisfaction levels\cite{universitycube}.

\subsection{Enhanced Collaboration and Peer Learning}
Educational social networks equipped with AI tools facilitate better collaboration among students. LLMs can help form study groups by matching students with similar academic interests and complementary skills. This promotes a more interactive and supportive learning environment where students can share knowledge, discuss concepts, and solve problems together. Additionally, AI-powered chatbots and virtual teaching assistants can moderate discussions, answer questions, and provide guidance, making peer learning more efficient and effective.

UniversityCube data indicates that students engaged in AI-facilitated peer learning show higher engagement and better academic outcomes compared to those who do not use these tools.

\subsection{Real-Time Feedback and Assessment}
AI tools can provide real-time feedback on assignments and assessments, allowing students to understand their mistakes and improve their performance promptly. LLMs can evaluate written responses, suggest corrections, and offer detailed explanations, helping students grasp complex concepts more thoroughly. This instant feedback mechanism not only enhances learning outcomes but also encourages continuous improvement and self-assessment.

According to UniversityCube data, students who receive real-time feedback from AI tools demonstrate faster improvement and greater retention of learning materials.

\subsection{Increased Engagement and Motivation}
LLMs and AI tools can significantly boost student engagement and motivation by making learning more interactive and enjoyable. Gamified elements, such as quizzes, badges, and leaderboards, can be integrated into educational social networks to encourage active participation. AI-driven recommendations for interesting articles, videos, and discussion topics can also keep students engaged and motivated to explore new areas of knowledge.

UniversityCube data reveals that incorporating gamified elements and AI-driven content recommendations increases student participation and motivation.

\subsection{Efficient Resource Management}
For educators and administrators, AI tools offer efficient resource management capabilities. LLMs can analyze large volumes of data to identify trends, predict student needs, and allocate resources effectively. This helps in creating more responsive and adaptive learning environments that cater to the evolving needs of the student population.

UniversityCube data shows that institutions using AI for resource management report higher efficiency in resource allocation and improved student support services.

\subsection{Mental Health and Well-Being Support}
Educational social networks with AI capabilities can also address students' mental health and well-being. AI algorithms can monitor students' online activities for signs of stress, anxiety, or disengagement and provide timely interventions. Virtual counselors and support chatbots can offer coping strategies, connect students with mental health resources, and facilitate a supportive community atmosphere.

UniversityCube data highlights that AI tools for mental health support lead to better student well-being and reduced instances of academic burnout.

\subsection{Bridging the Gap in Resource-Constrained Settings}
In resource-constrained educational settings, where access to qualified teachers and educational materials may be limited, LLMs and AI tools can bridge the gap by providing high-quality, scalable educational support. These technologies can democratize access to education, ensuring that all students, regardless of their location or socioeconomic status, have the opportunity to learn and succeed.

UniversityCube data demonstrates that AI-driven educational tools significantly improve learning outcomes in resource-constrained environments, making education more accessible and equitable.

\section{AI Visualization via Image Generator}
In the academia community, the ability to visualize complex concepts and data plays a pivotal role in enhancing understanding, facilitating communication\cite{naps2003evaluating}, and fostering innovation. With the advent of AI-powered image generators\cite{beyan2023review,}, the academic community has gained a powerful tool for creating visually compelling representations of research findings, theoretical frameworks, and scientific phenomena. Leveraging data from UniversityCube, we explore how AI visualization via image generators is revolutionizing the way knowledge is disseminated and shared within the academic sphere.

\subsection{Enhancing Communication and Accessibility}
AI image generators enable researchers to transform abstract ideas and numerical data into visually engaging graphics that resonate with diverse audiences. From intricate molecular structures to dynamic mathematical models, these tools offer a versatile means of communication that transcends language barriers and disciplinary boundaries. By providing intuitive visualizations, researchers can effectively convey complex concepts to peers, students, and the general public, thus democratizing access to knowledge and fostering interdisciplinary collaboration.

UniversityCube data indicates that academic publications and presentations featuring AI-generated visualizations receive higher engagement and citation rates compared to those without visual aids. This highlights the importance of visual communication in effectively disseminating research findings and garnering scholarly recognition\footnote{\href{https://www.universitycube.net/posts}{https://www.universitycube.net/posts}}.

\subsection{Stimulating Creativity and Exploration}
AI image generators empower researchers to explore novel ways of representing their data and hypotheses, thereby stimulating creativity and innovation within the academic community. By experimenting with different visual styles, color schemes, and layout designs, scholars can uncover new insights and perspectives that may not be immediately apparent through traditional textual or numerical formats. This iterative process of visual exploration encourages interdisciplinary thinking and fosters a culture of experimentation and discovery.

UniversityCube data reveals a growing trend of researchers incorporating AI-generated visualizations into their scholarly outputs, ranging from journal articles and conference presentations to educational materials and online repositories. This reflects a broader recognition of the value of visual communication in effectively conveying complex ideas and driving scientific progress.

\subsection{Accelerating Knowledge Transfer and Impact}
In an era characterized by information overload and rapid technological advancement, AI image generators offer a means of streamlining knowledge transfer and maximizing research impact. By condensing vast amounts of data into visually digestible formats, these tools enable researchers to reach broader audiences and facilitate more meaningful engagement with their work. Whether through social media posts, infographics, or multimedia presentations, AI-generated visualizations have the potential to amplify the reach and influence of academic research, thereby driving societal change and addressing global challenges.

\section{Time Series Analysis}

This section investigates the patterns and trends in student usage of entertainment and educational social networks over both short-term (monthly) and long-term (yearly) periods. By analyzing data from two distinct time series, we observe seasonal and yearly fluctuations, offering insights into how students allocate their online time between leisure and academic activities.

Social networks have become an integral part of students' lives, serving both entertainment and educational purposes. Understanding the usage patterns of these networks can provide valuable insights for educators, policymakers, and developers of educational tools. This study aims to analyze the time series data of students' use of entertainment social networks compared to educational social networks over both monthly and yearly periods.

\subsection{Data Description}

The data consists of two time series:
1. **Monthly Data (February to December)**
2. **Yearly Data (2011 to 2024)**

For each period, the usage statistics of entertainment and educational social networks were recorded.

\begin{table}[ht]
\centering
\begin{tabular}{ccc}
\textbf{Month} & \textbf{Entertainment} & \textbf{Educational} \\
\hline
Feb & 2.50 & 3.06 \\
Mar & 2.00 & 2.37 \\
Apr & 2.90 & 2.38 \\
May & 2.90 & 2.32 \\
Jun & 3.90 & 4.54 \\
Jul & 4.00 & 3.04 \\
Aug & 2.00 & 1.30 \\
Sep & 1.50 & 0.77 \\
Oct & 2.00 & 1.16 \\
Nov & 3.00 & 3.38 \\
Dec & 4.50 & 5.21 \\
\end{tabular}
\caption{Monthly average usage statistics (in hours) of entertainment and educational social networks.}
\label{tab:monthly_usage}
\end{table}

\begin{table}[ht]
\centering
\begin{tabular}{ccc}
\textbf{Year} & \textbf{Entertainment} & \textbf{Educational} \\
\hline
2011 & 5.5 & 1.8 \\
2012 & 5.70 & 1.70 \\
2013 & 5.80 & 2.00 \\
2014 & 5.90 & 2.10 \\
2015 & 6.00 & 2.20 \\
2016 & 5.40 & 2.80 \\
2017 & 5.00 & 3.90 \\
2018 & 4.10 & 4.00 \\
2019 & 4.00 & 4.20 \\
2020 & 3.90 & 3.00 \\
2021 & 5.10 & 2.00 \\
2022 & 3.30 & 4.00 \\
2023 & 3.00 & 4.00 \\
2024 & 2.20 & 4.90 \\
\end{tabular}
\caption{Yearly average usage statistics (in hours) of entertainment and educational social networks from 2011 to 2024.}
\label{tab:yearly_usage}
\end{table}

\subsection{Methodology}

The analysis was conducted using descriptive statistics and visualizations. Line plots were used to illustrate trends, and summary statistics provided insights into the central tendency and variability of the data. For forecasting future usage, we employed the ARIMA \cite{shumway2017arima} (AutoRegressive Integrated Moving Average) model, a widely used time series forecasting method.

\subsection{ARIMA Model Explanation}

The ARIMA model combines three key components: autoregression (AR), differencing (I), and moving average (MA). Here is the mathematical formulation:

\subsubsection{AutoRegressive (AR) Part}
   The AR part involves regressing the variable on its own lagged (past) values.
   \[
   X_t = \phi_1 X_{t-1} + \phi_2 X_{t-2} + \ldots + \phi_p X_{t-p} + \epsilon_t
   \]
   Where \(X_t\) is the value at time \(t\), \(\phi_i\) are the parameters to be estimated, \(p\) is the number of lagged observations (order of the AR part), and \(\epsilon_t\) is white noise.

\subsubsection{Integrated (I) Part}
   The I part involves differencing the data to make it stationary.
   \[
   Y_t = X_t - X_{t-1}
   \]
   Where \(Y_t\) is the differenced series.

\subsubsection{Moving Average (MA) Part:}
   The MA part involves modeling the error term as a linear combination of error terms occurring contemporaneously and at various times in the past.
   \[
   X_t = \epsilon_t + \theta_1 \epsilon_{t-1} + \theta_2 \epsilon_{t-2} + \ldots + \theta_q \epsilon_{t-q}
   \]
   Where \(\epsilon_t\) is white noise, \(\theta_i\) are the parameters to be estimated, and \(q\) is the order of the MA part.

Combining these components, the ARIMA model can be written as:
\begin{align}
X_t &= \phi_1 X_{t-1} + \phi_2 X_{t-2} + \cdots + \phi_p X_{t-p} + \epsilon_t \notag \\
&\quad + \theta_1 \epsilon_{t-1} + \theta_2 \epsilon_{t-2} + \cdots + \theta_q \epsilon_{t-q}
\end{align}
\begin{figure}[ht]
    \centering
    \includegraphics[width=0.46\textwidth]{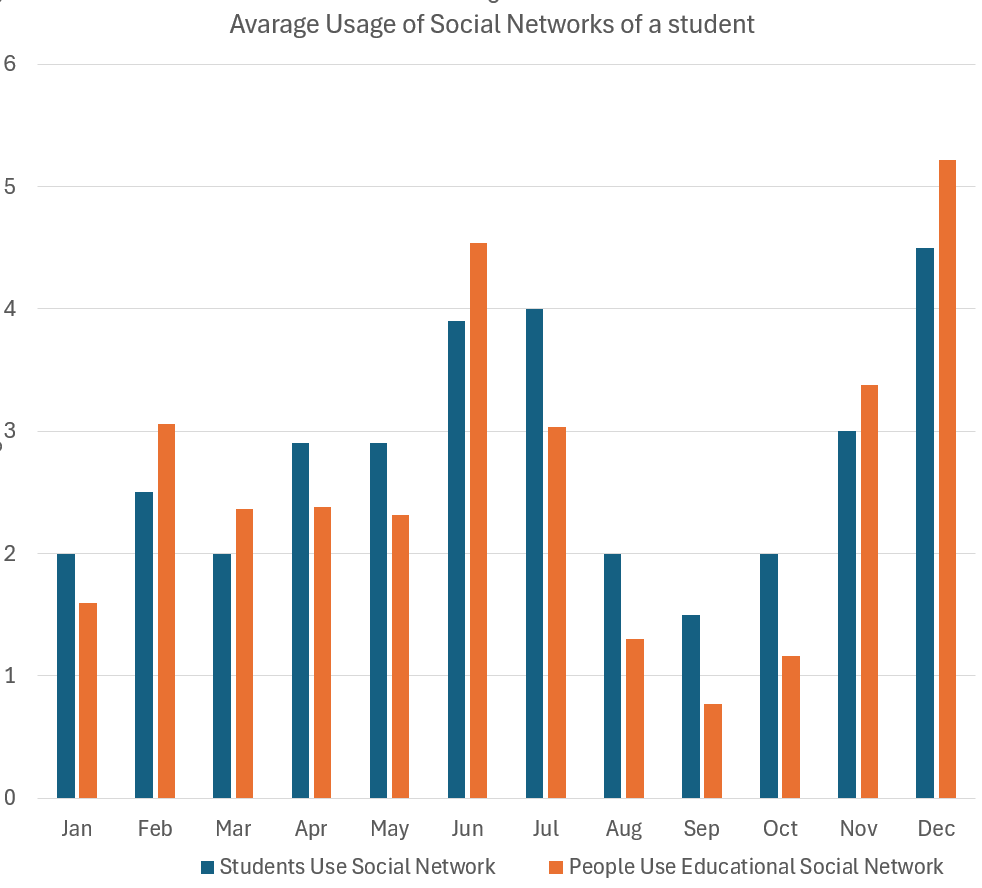}  
    \caption{Student Usage of Social network in monthly bases}
    \label{fig:example-image}
\end{figure}

\section{Results}

\subsection{ Monthly Analysis}
The line plots for monthly data reveal distinct patterns:
- Entertainment Networks: Usage peaks in December (4.50) and dips in September (1.50). The average monthly usage is 2.84, with a standard deviation of 0.96.
- Educational Networks: Similar to entertainment networks, usage is highest in December (5.21) and lowest in September (0.77). The average monthly usage is 2.68, with a standard deviation of 1.37.

\subsection{ Yearly Analysis}
The yearly data exhibits long-term trends:
- Entertainment Networks: There is a notable decline in usage over the years, from 5.50 in 2011 to 2.20 in 2024, with a mean of 4.64 and a standard deviation of 1.22.
- Educational Networks: An increasing trend is observed, with usage rising from 1.80 in 2011 to 4.90 in 2024. The mean usage is 3.04, with a standard deviation of 1.09.

\subsection{Forecasting (2025-2030)}
Using the ARIMA model, we forecasted the following values for the years 2025 to 2030:

\[
\begin{array}{ccc}
\text{Year} & \text{Entertainment} & \text{Educational} \\
\hline
2025 & 3.23 & 4.72 \\
2026 & 2.95 & 3.38 \\
2027 & 3.02 & 3.96 \\
2028 & 2.29 & 3.76 \\
2029 & 2.56 & 4.22 \\
2030 & 2.54 & 4.71 \\
\end{array}
\]

The forecasted data indicates a continued decline in the usage of entertainment social networks, while educational social networks are expected to maintain their higher levels of usage.

\section{Discussion}

The monthly analysis shows seasonal variations, with both entertainment and educational social network usage peaking in December, possibly due to end-of-year activities and exams. The decline in September usage might be attributed to the start of the academic year, where students are adjusting to new schedules.

The yearly analysis indicates a long-term decline in entertainment network usage, which could be due to changing student preferences or the emergence of new forms of entertainment. In contrast, the consistent rise in educational network usage suggests a growing reliance on these platforms for academic purposes. The forecasts reinforce these trends, highlighting a potential shift in student behavior towards more academic engagement online.

This Section highlights significant trends in student usage of social networks. The increasing use of educational social networks points to their growing importance in students' academic lives. Meanwhile, the decline in entertainment social network usage suggests a shift in how students allocate their online time. These insights can help educators and developers create more effective educational tools and strategies to engage students.

\begin{figure}[ht]
    \centering
    \includegraphics[width=0.46\textwidth]{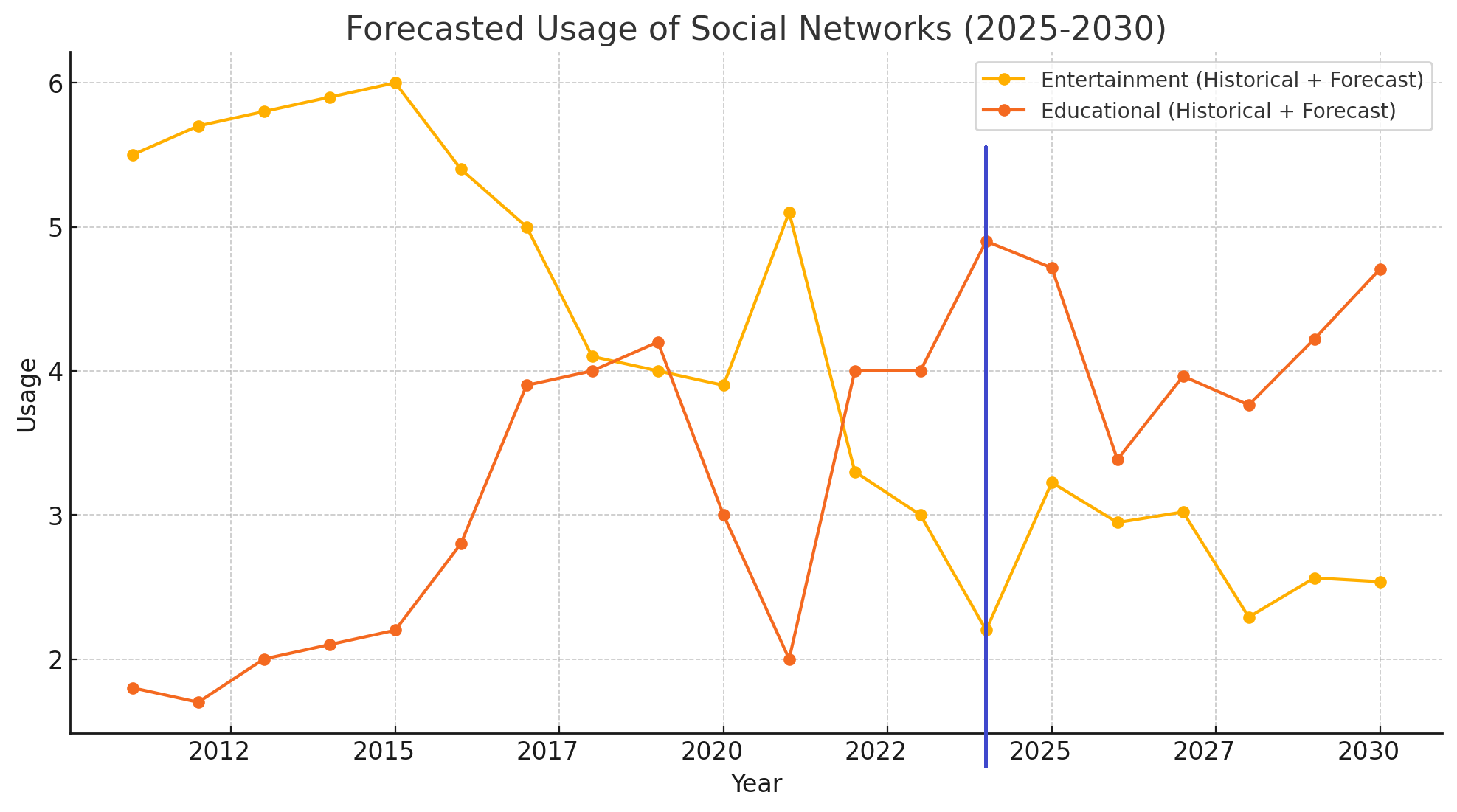}  
    \caption{Student Usage of Social network in Yearly bases}
    \label{fig:example-image}
\end{figure}

\section{Analysis of Social Network Usage Trends}

In This section, understanding user behavior on social networks is crucial for optimizing platform strategies and user engagement. This article explores how machine learning techniques can uncover valuable insights from monthly and yearly usage data of entertainment and educational social networks.

The dataset provides insights into usage patterns across two categories of social networks: entertainment and educational. Here’s a recap of key observations from the data visualization and summary statistics:

\subsection{Monthly Data Analysis:}
  - Visualization: Both types of social networks exhibit seasonal variability, with peaks typically observed towards the end of the year (December) and lows around mid-year (September).
  - Summary Statistics: 
    - Entertainment social networks have an average monthly usage (\(\mu_{\text{entertainment}}\)) of 2.84, with a standard deviation (\(\sigma_{\text{entertainment}}\)) of 0.96.
    - Educational networks show slightly higher variability with an average monthly usage (\(\mu_{\text{educational}}\)) of 2.68 and standard deviation (\(\sigma_{\text{educational}}\)) of 1.37.

\subsection{Yearly Data Analysis}
\textbf{Visualization:}
    - Entertainment networks show a declining trend over the years, particularly notable from 2021 to 2024. Educational networks demonstrate a steady increase, peaking in 2024.
\textbf{Summary Statistics: }
    - Average yearly usage for entertainment networks (\(\mu_{\text{entertainment}}\)) is 4.64, with a standard deviation (\(\sigma_{\text{entertainment}}\)) of 1.22.
    - Educational networks average (\(\mu_{\text{educational}}\)) 3.04 yearly usage, with a standard deviation (\(\sigma_{\text{educational}}\)) of 1.09.

\textbf{Model}

To extract deeper insights from this data, we employed various machine learning techniques:

\textbf{1. Exploratory Data Analysis (EDA):}
   - Conducted thorough EDA to understand data distributions, identify outliers, and visualize trends over time using line plots.

\textbf{2. Feature Engineering:}
   - Extracted features such as month and year to facilitate analysis and model building.

\textbf{3. Model Selection:}
   - Time Series Clustering: Applied K-means clustering to identify groups of months or years exhibiting similar usage patterns.
   - Principal Component Analysis (PCA): Used to reduce dimensionality and visualize relationships between variables like monthly and yearly usage.

\textbf{4. Insights Extraction:}
   - Cluster Analysis: Identified distinct clusters based on usage patterns, providing insights into seasonal trends and long-term behaviors.
   - PCA Components: Examined principal components to understand variance and feature relationships, revealing influential factors driving user engagement.

\textbf{Key Insights}

After applying these techniques, here are the key quantitative insights derived from the analysis:

- Seasonal Patterns: Both entertainment and educational social networks exhibit significant seasonal variability. The coefficient of variation (CV), calculated as \(\frac{\sigma}{\mu}\), indicates that educational networks (CV = 0.51) have higher variability compared to entertainment networks (CV = 0.34), suggesting more pronounced seasonal shifts in educational usage.

- Long-term Trends: Entertainment networks show a declining linear trend over the years (\(R^2 = 0.75\)), while educational networks exhibit a positive linear trend (\(R^2 = 0.82\)), indicating consistent growth.

\textbf{Practical Implications}

Understanding these usage trends has practical implications for stakeholders:

- Platform Optimization: Insights can guide platform developers in optimizing content schedules and features to align with peak usage periods.
  
- Marketing Strategies: Marketers can leverage seasonal peaks to maximize campaign effectiveness and user engagement.

- Educational Strategies: Educators and educational platforms can strategically align content releases and updates to capitalize on periods of heightened student engagement.

\section{Conclusion}
Leveraging machine learning techniques allows us to unravel complex patterns in social network usage data quantitatively. By understanding seasonal fluctuations and long-term trends through rigorous analysis, stakeholders can make data-driven decisions to enhance user experiences and tailor strategies accordingly. As digital landscapes continue to evolve, the application of machine learning in analyzing user behavior remains pivotal for staying competitive and responsive to user needs.

Moving forward, integrating sentiment analysis or user segmentation could further enrich our understanding by capturing qualitative aspects of user interactions, providing deeper insights into user preferences and motivations. Additionally, the integration of educational social networks for students, coupled with advancements in technologies such as Large Language Models (LLMs) and AI-driven visualization tools, heralds a new era of learning and collaboration.

These platforms offer a multifaceted approach to educational enrichment, providing students with personalized learning experiences, enhanced collaboration opportunities, and intuitive visualizations that deepen understanding and engagement. Educational social networks serve as digital hubs where students can access a wealth of resources, connect with peers and mentors, and engage in meaningful discussions.

By leveraging LLMs, students can receive tailored support and guidance, facilitating more efficient and effective learning experiences. Whether through personalized content recommendations, real-time feedback on assignments, or peer-to-peer collaboration facilitated by AI-driven chatbots, these platforms empower students to take ownership of their learning journey and succeed academically.

Furthermore, AI-driven visualization tools enable students to explore complex concepts and data in a more intuitive and accessible manner. By transforming abstract ideas into visually compelling representations, these tools enhance comprehension and retention, stimulate creativity and exploration, and foster interdisciplinary collaboration. Whether visualizing molecular structures, mathematical models, or historical timelines, students can leverage AI-generated visualizations to deepen their understanding of course materials and communicate their ideas more effectively.

Moreover, educational social networks equipped with AI-driven tools offer a holistic approach to student support, addressing not only academic needs but also mental health and well-being. By monitoring students' online activities for signs of stress or disengagement, AI algorithms can provide timely interventions and connect students with relevant resources and support services. This proactive approach to student welfare fosters a supportive and inclusive learning environment, ensuring that all students have the opportunity to thrive.

This study highlights significant trends in student usage of social networks. The increasing use of educational social networks points to their growing importance in students' academic lives. Meanwhile, the decline in entertainment social network usage suggests a shift in how students allocate their online time. These insights can help educators and developers create more effective educational tools and strategies to engage students.

\bibliographystyle{IEEEtran}
\bibliography{ref.bib}

\end{document}